 \documentclass[11pt]{article}
 \usepackage[top=1in,bottom=1in,left=1.5in,right=1.5in]{geometry}
 \usepackage{amsmath,amssymb}
 \usepackage{latexsym}% defines $\Box$ for LaTeX2e
 \usepackage[dvips]{pstricks} % PSTricks
 \usepackage[dvips]{graphicx}
 \newcommand{\N}{\mathbb{N}}
 \newcommand{\R}{\mathbb{R}}

 \renewcommand{\d}{\mathbf{d}}

 \newcommand{\x}{\mathbf{x}}

 \newcommand{\1}{\mathbf{1}}

 \newcommand{\cA}{\mathcal{A}}

 \newcommand{\cP}{\mathcal{P}}

 \newcommand{\lan}{\langle}
 \newcommand{\ran}{\rangle}

 \newcommand{\an}[1]{\lan#1\ran}
 \newcommand{\hs}{\hspace*{\parindent}}
 \newcommand{\proof}{\hs \textbf{Proof.\ }}
 \newcommand{\tr}{\mathop{\mathrm{tr}}\nolimits}

 \newcommand{\trans}{^\top}
 \newcommand{\qed}{\hspace*{\fill} $\Box$\\}

 \newtheorem{theo}{\bfseries \hs Theorem}[section]
 
 \newtheorem{prop}[theo]{\bfseries \hs Proposition}
 \newtheorem{lemma}[theo]{\bfseries \hs Lemma}
 \newtheorem{corol}[theo]{\bfseries \hs Corollary}

 \numberwithin{equation}{section}
\begin{document}

 \title{On the graph isomorphism problem}

 \author{Shmuel Friedland
  \thanks{Department of Mathematics, Statistics, and Computer Science,
  University of Illinois at Chicago
  Chicago, Illinois 60607-7045, USA, \texttt{E-mail: friedlan@uic.edu}}
  \thanks{Visiting Professor,
  Berlin Mathematical School,
  Institut f\"{u}r Mathematik,
  Technische Universit\"{a}t Berlin,
  Strasse des 17. Juni 136,
  D-10623 Berlin, Germany}}

 \date{January 9, 2008}

 \maketitle

 \begin{abstract}
 We relate the graph isomorphism problem to the solvability
 of certain systems of linear equations and linear inequalities.
 The number of these equations and inequalities is related to the
 complexity of the graphs isomorphism and subgraph isomorphim problems.
 \\



 \end{abstract}

%-------------------------------------------------------------------------------

\section{Introduction}

 Let $G_1=(V,E_1),G_2=(V,E_2)$ be two simple undirected graphs,
 where $V$ is the set of vertices of cardinality $n$ and
 $E_1,E_2\subset V\times V$ the set of edges. $G_1$ and $G_2$
 are called \emph{isomorphic} if there exists a bijection $\sigma:V\to V$
 which induces the corresponding bijection $\tilde
 \sigma:E_1\to  E_2$.
 The graph isomorphism problem, abbreviated here as \emph{GIP},
 is the problem of determination if $G_1$ and $G_2$ are isomorphic.
 Clearly the \emph{GIP} in the class \emph{NP}.
 It is one of a very small number of problems whose complexity
 is unknown \cite{GJ, KST}.  For certain graphs it is known that the complexity
 of \emph{GIP} is polynomial \cite{BGM,Bod,FM,HW,Luk,Mil}.

 Let $G_3=(W,E_3)$, where $\#W=m\le n$.
 $G_3$ is called isomorphic to a subgraph of $G_2$ if there
 exits an injection $\tau: V_3 \to V_2$ which induces an
 injection $\tilde \tau:E_3\to E_2$.
 The subgraph isomorphism, abbreviated here as \emph{SGIP},
 is the problem of determination
 if $G_3$ is isomorphic to a subgraph of $G_2$.
 It is well known that SGIP is \emph{NP-Complete} \cite{GJ}.

 In the previous versions of this paper we related the graph
 isomorphism problem to the solvability
 of certain systems of linear equations and linear inequalities.
 It was pointed out to me by N. Alon and L. Babai, that my
 approach relates in a similar way the SGIP to the solvability
 of certain systems of linear equations and linear inequalities.
 Hence $f(n)$, the number of these linear equalities and inequalities
 for $V=n$, is probably exponential in $n$.
 Thus, the suggested approach in this paper does not seem to be
 the right approach to determine the complexity of the GIP.
 Nevertheless, in this paper we summarize the main ideas and
 results of this approach.  It seems that our approach is
 related to the ideas and results discussed in \cite{Onn}.

 Let $\Omega_n\subset \R_+^{n\times n}$ be the convex set of
 $n\times n$ doubly stochastic matrices.
 In this paper we relate the complexity of the \emph{GIP} to
 the minimal number of supporting hyperplanes determining a
 certain convex polytope $\Psi_{n,n}\subset
 \Omega_{n^2}$.  More precisely, two graph are isomorphic if
 certain system of $n^2$ hyperplanes intersect $\Psi_{n,n}$.
 More general, if the corresponding system $n^2$ half spaces
 intersect $\Psi_{n,n}$ then $G_3$ is isomorphic
 to a subgraph of $G_2$.
 Hence the minimal number of supporting hyperplanes defining $\Psi_{n,n}$,
 denoted by $f(n)$,  is closely related to the complexity of
 $SGIP$.  We give a larger polytope $\Phi_{n,n}$, characterized by
 $(4n-1)n^2$ linear equations in $n^4$ nonnegative variables
 satisfying
 \begin{equation}\label{minclusion}
 \Psi_{n,n}\subset \Phi_{n,n}\subset \Omega_{n^2}.
 \end{equation}

 \emph{In the first version of this paper we erroneously
 claimed that} $\Phi_{n,n}=\Psi_{n,n}$.  The error in my proof
 was pointed out to me by Babai, Melkebeek, Rosenberg and
 Vavasis.  The inequality $\Psi_{n,n}\subsetneqq \Phi_{n,n}$
 for $n\ge 4$ is implied by the example of J. Rosenberg.

 Thus if two graphs are isomorphic then
 certain system of $n^2$ hyperplanes intersect $\Phi_{n,n}$.
 This of course yields a necessary conditions for GIP and SGIP.

 We now outline the main ideas of the paper.  Let $A,B$ be
 $n\times n$ adjacency matrices of $G_1,G_2$.  So $A,B$ are
 $0-1$ symmetric matrices with zero diagonal.  It is enough
 to consider the case where $A$ and $B$ have the same number of
 $1$'s.  Let $\cP_n$ be the set of $n\times n$ permutation
 matrices.  Then $G_1$ and $G_2$ are isomorphic if and only if
 $PAP\trans=B$ for some $P\in\cP_n$.  It is easy to see that
 this condition is equivalent

 \begin{equation}\label{pqcond}
 P(A+2I_n)Q\trans = B+2I_n   \textrm{ for some } P,Q\in
 \cP_n,
 \end{equation}
 where $I_n$ is the $n\times n$ identity matrix.

 For $C,D\in\R^{n\times n}$ denote by $C\otimes D\in \R^{n^2\times n^2}$ the
 Kronecker product, see \S2.  Let $\cP_n\otimes
 \cP_n:=\{P\otimes Q,\; P,Q\in\cP_n\}$.  Denote by
 $\Psi_{n,n}\subset \R_+^{n^2\times n^2}$ the convex set
 spanned by $\cP_n\otimes\cP_n$.  $\Psi_{n,n}$ is a subset of
 $n^2\times n^2$
 doubly stochastic matrices.  Then the condition (\ref{pqcond})
 implies the solvability of the system of $n^2$ equations of
 the form $Z(\widehat{A+2I_n})=\widehat{B+2I_n}$ for some $Z\in
 \Psi_{n,n}$.
 Here $\widehat{B+2I_n}\in\R^{n^2}$ is a column vector composed
 of the columns of $B+2I_n$.
 Vice versa, the solvability of $Z(\widehat{A+2I_n})=\widehat{B+2I_n}$ for some $Z\in
 \Psi_{n,n}$ implies (\ref{pqcond}).
 The ellipsoid algorithm in linear programming
 \cite{Kha,Kar} yields that the existence
 a solution to this system of equations is determined in polynomial time in $\max(f(n),n)$.
 Similarly, for the SGIP one needs to consider the the
 solvability of $Z(\widehat{C+2^{n^2}I_n})\le \widehat{B+2^{n^2}I_n}$ for some $Z\in
 \Psi_{n,n}$, where $C$ is the adjacency matrix of the graph
 $\tilde G_3=(V,E_3)$ obtained from $G_3$ by appending $n-m$
 isolated vertices.

 We now survey briefly the contents of this paper.
 In \S2 we introduce the needed concepts from linear algebra
 to give the characterization of $\Phi_{n,n}$ in terms of
 $(4n-2)n^2$ linear equations in $n^4$ nonnegative variables.
 This is done for the general set $\Phi_{m,n}$, which contains $\Psi_{m,n}$,
 the convex hull of $\cP_m\otimes \cP_n$.
 \S3 discusses the permutational similarity of $A,B\in
 \R^{n\times n}$ and permutational equivalence of $A,B\in
 \R^{n\times m}$.  We show the second main result that
 the permutational similarity and equivalence is equivalent to
 solvability of the corresponding system of equations discussed
 above.  In \S4 we deduce the complexity results claimed in
 this paper.

 This paper generated a lot of interest. I would like to thank
 all the people who sent their comments to me.

\section{Tensor products of doubly stochastic matrices}

 For $m\in\N$ denote $\an{m}:=\{1,\ldots,m\}$.
 For $\cA\subset \R$ denote by $\cA^{m\times n}$ the set of
 $m\times n$ matrices $A=[a_{ij}]_{i,j=1}^{m,n}$ such that each
 $a_{ij}\in\cA$. Recall that $A=[a_{ij}]\in \R_+^{m\times m}$
 is called doubly stochastic if
 \begin{equation}\label{defds}
 \sum_{j=1}^m a_{ij}=\sum_{j=1}^m a_{ji}=1, \quad i=1,\ldots,m.
 \end{equation}
 Since the sum of all rows of $A$ is equal to the sum of all
 columns of $A$ it follows that at most $2m-1$ of above equations
 are linearly independent.  It is well known that any $2m-1$ of
 the above equation are linearly independent.
 Let $\1:=(1,\ldots,1)\trans\in\R_+^m$.  Note that
 $A=[a_{ij}]\in\R^{m\times m}$ satisfies (\ref{defds}) if and only if and
 $A\1= A\trans \1=\1$.

 Denote by $\Omega_m$ the set of doubly stochastic matrices.
 Clearly, $\Omega_m$ is a convex compact set.  Birkhoff theorem
 claims that the set of the extreme points of $\Omega_m$ is the
 set of permutations matrices $\cP_m\subset \{0,1\}^{m\times
 m}$.

 \begin{lemma}\label{scaldsm}
 Denote by $\Lambda_m\subset \R_+^{m\times m}$ the set of
 nonnegative matrices satisfying the conditions $A\1=A\trans
 \1=a\1$ for some $a\ge 0$ depending on $A$.  Then $\Lambda_m$ is a
 multiplicative cone:
 $$\Lambda_m+\Lambda_m=\Lambda_m,\; a\Lambda_m\subset \Lambda_m
 \textrm{ for all } a\ge 0,\; \Lambda_m \cdot
 \Lambda_m=\Lambda_m.
 $$
 Furthermore,  $A=[a_{ij}]\in \R_+^{m\times m}$ is in $\Lambda_m$
 if and only if the following $2(m-1)$ equalities hold.
 \begin{equation}\label{scaldsmcon}
 \sum_{j=1}^m a_{ij}=\sum_{j=1}^m a_{ji}=\sum_{j=1}^m a_{1j} \textrm{ for
 } i=2,\ldots,m.
 \end{equation}
 \end{lemma}

 \proof  The fact that $\Lambda_m$ is a cone is
 straightforward.  Since $I_m\in \Lambda_m$ we deduce the
 equality $\Lambda_m\cdot\Lambda_m=\Lambda_m$.
 Observe next that the conditions (\ref{scaldsmcon}) imply that
 $A\1=a\1$, where $a$ is the sum of the elements in the first
 row.  Also the sum of the elements in each column except the first is equal
 to $a$.  Since the sum of all elements of $A$ is $ma$ it
 follows that the sum of the elements in the first column is
 also $a$, i.e $A\trans \1=a\1$.  \qed

 For $A=[a_{ij}]\in \R^{m\times m},B=[b_{k l}]\in \R^{n\times
 n}$ denote by $A\otimes B\in \R^{mn \times mn}$ the tensor
 product of $A$ and $B$.  The rows and columns of
 $A\otimes B$ are indexed by double indices $(i,k)$ and
 $(j,l)$, where $i,j=1,\ldots m, \;k,l=1,\ldots,n$.
 Thus
 \begin{eqnarray}\label{repAtenB}
 A\otimes B=[c_{(i,k)(j,l)}]\in \R^{mn\times mn},\\
 \textrm{where }
 c_{(i,k)(j,l)}=a_{ij}b_{kl}  \textrm{ for }
 i,j=1,\ldots,m,\;k,l=1,\ldots,n.\nonumber
 \end{eqnarray}

 If we arrange the indices
 $(i,k)$ in the lexicographic order then $A\otimes
 B$ has the following block matrix form called the
 \emph{Kronecker product}

 \begin{equation}\label{kronpr}
 A\otimes B=\left[\begin{array}{rrrr}a_{11}B&a_{12}B&\ldots&
 a_{1m}B\\ a_{21}B&a_{22}B&\ldots&a_{2m}B\\
 \vdots&\vdots&\ldots&\vdots\\
 a_{m1}B&a_{m2}B&\ldots&a_{mm}B\end{array}\right].
 \end{equation}
 For simplicity of the exposition we will identify $A\otimes B$
 with the block matrix (\ref{kronpr}) unless stated otherwise.
 Note that any other ordering of $\an{m}\times\an{n}$
 induces a different representation of $A\otimes B$ as $C\in
 \R^{mn\times mn}$, where $C=P (A\otimes B) P\trans$ for some
 permutation matrix $P\in \cP_{mn}$.

 Recall that $A\otimes B$ is bilinear in $A$ and $B$.
 Furthermore

 \begin{equation}\label{prodtenprod}
 (A\otimes B)(C\otimes D)=(AC)\otimes (BD) \textrm{ for all }
 A,C\in \R^{m\times m},\; B,D\in \R^{n\times n}.
 \end{equation}

 \begin{prop}\label{tenproddsm} Let $A\in \Omega_m, B\in
 \Omega_n$.  Then $A\otimes B\in \Omega_{mn}$.
 \end{prop}

 \proof Clearly $A\otimes B$ is a nonnegative matrix.
 Assume the representation (\ref{repAtenB}).
 Then
 \begin{eqnarray*}
 \sum_{j,l=1}^{m,n} c_{(i,k)(j,l)}=\sum_{j,l=1}^{m,n}
 a_{ij}b_{kl}=(\sum_{j=1}^m a_{ij})(\sum_{l=1}^n b_{kl})=1
 \cdot 1=1,\\
 \sum_{j,l=1}^{m,n} c_{(j,l)(i,k)}=\sum_{j,l=1}^{m,n}
 a_{ji}b_{lk}=(\sum_{j=1}^m a_{ji})(\sum_{l=1}^n b_{lk})=1
 \cdot 1=1.
 \end{eqnarray*}
 \qed

 \begin{lemma}\label{conhtproddsm}
 Denote by $\Psi_{m,n}\subset \Omega_{mn}$ the convex hull
 spanned by $\Omega_m\otimes \Omega_n$, i.e. all doubly
 stochastic matrices of the form $A\otimes B$, where
 $A\in\Omega_m,\;B\in \Omega_n$.  Then the extreme points
 of $\Psi_{m,n}$ is the set $\cP_m\otimes \cP_n$, i.e. each
 extreme point is of the form $P\otimes Q$, where $P\in\cP_m,
 Q\in \cP_n$.
 \end{lemma}

 \proof  Use Birkhoff's theorem and the bilinearity of $A\otimes B$
 to deduce that $\Psi_{m,n}$ is spanned by $\cP_m\otimes
 \cP_n$.  Clearly $\cP_m\otimes \cP_n\subset \cP_{mn}$.
 Since Birkhoff's theorem implies that $\cP_{mn}$ are extreme
 points of $\Omega_{mn}$ it follows that $\cP_m\otimes \cP_n\subset \cP_{mn}$
 are convexly independent.  \qed

 \begin{theo}\label{charpsimn} Let $\Phi_{m,n}$ be the convex
 set of $mn\times mn$ nonnegative
 matrices characterized by $2mn+(2n-2)m^2+(2m-2)n^2$ linear equations of the
 following form.  View $C\in \R^{mn\times mn}$ as a matrix with
 entries $c_{(i,k)(j,l)}$ where
 $i,j=1,\ldots,m,\;k,l=1,\ldots,n$.  Then $C\in \R_+^{mn\times
 mn}$ belongs to $\Phi_{m,n}$ if the following
 equalities hold.

 \begin{eqnarray}
 \sum_{j,l=1}^{m,n} c_{(i,k),(j,l)}= \sum_{j,l=1}^{m,n}
 c_{(j,l)(i,k)}=1,\; i=1,\ldots,m,\;k=1,\ldots,n,
 \label{dscon}\\
 \sum_{j=1}^m c_{(i,k)(j,l)}=\sum_{j=1}^m c_{(1,k)(j,l)},\;
 \sum_{j=1}^m c_{(j,k)(i,l)}=\sum_{j=1}^m c_{(1,k)(j,l)}
 \label{klcond}\\
 \textrm{where } i=2,\ldots,m \textrm{ and } k,l=1,\ldots,n,
 \nonumber\\
 \sum_{l=1}^n c_{(i,k)(j,l)}=\sum_{l=1}^n c_{(i,1)(j,l)},\;
 \sum_{l=1}^n c_{(i,l)(j,k)}=\sum_{l=1}^n c_{(i,1)(j,l)}
 \label{ijcond}\\
 \textrm{where } k=2,\ldots,n \textrm{ and } i,j=1,\ldots,m.
 \nonumber
 \end{eqnarray}

 Furthermore
 \begin{equation}\label{minclusionmn}
 \Psi_{m,n}\subset \Phi_{m,n}\subset\Omega_{mn}
 \end{equation}
 \end{theo}

 \proof The conditions (\ref{dscon}) state that $C\in
 \Omega_{mn}$.  We now show the conditions $\Psi_{m,n}\subseteq\Phi_{m,n}$.
 Let $A\in \Omega_m, B\in \Omega_n$ and consider the Kronecker
 product (\ref{kronpr}).  Then for $i,j\in \an{m}$, the $(i,j)$
 block of $A\otimes B$ is $a_{ij}B\in \Lambda_n$.
 Since $\Lambda_n$ is a cone, it follows that for any
 $C\in \Psi_{m,n}$, having the block form $C=[C_{ij}],\; C_{ij}\in
 \R_+^{n\times n}, \;i,j\in \an{m}$, each $C_{ij}\in \Lambda_n$.
 Lemma \ref{scaldsm} yields the conditions for each $i,j\in \an{m}$.
 Since $A\otimes B=
 P(B\otimes A)P\trans$ we also deduce the conditions (\ref{klcond})
 for each $k,l\in \an{n}$.  \qed

 \begin{lemma}\label{eq22case} $\Psi_{2,2}=\Phi_{2,2}$.
 \end{lemma}
 \proof Let $D=[d_{pq}]_{p,q=1}^4\in \Phi_{2,2}$.  Since
 $$F_{11}:=\left[\begin{array}{rr}
 d_{11}&d_{12}\\d_{21}&d_{22}\end{array}\right], F_{12}:=\left[\begin{array}{rr}
 d_{13}&d_{14}\\d_{23}&d_{24}\end{array}\right]\in \Lambda_2$$
 it follows that
 $$d_{11}=d_{22}=a,\; d_{12}=d_{21}=b,\;
 d_{13}=d_{24}=c,\;d_{14}=d_{23}=d.$$
 Since
 $$G_{11}:=\left[\begin{array}{rr}
 d_{11}&d_{13}\\d_{31}&d_{33}\end{array}\right], G_{12}:=\left[\begin{array}{rr}
 d_{12}&d_{14}\\d_{32}&d_{34}\end{array}\right]\in \Lambda_2$$
 it follows that
 $$d_{31}=c,\; d_{32}=d,\;
 d_{33}=a,\;d_{34}=b.$$
 Since
 $$F_{21}:=\left[\begin{array}{rr}
 d_{31}&d_{32}\\d_{41}&d_{42}\end{array}\right], F_{22}:=\left[\begin{array}{rr}
 d_{33}&d_{34}\\d_{43}&d_{44}\end{array}\right]\in \Lambda_2$$
 it follows that
  $$d_{41}=d,\; d_{42}=c,\;
 d_{43}=b,\;d_{44}=b.$$
 So $a,b,c,d\ge 0$ and $a+b+c+d=1$.  This set has $4$ extreme
 points which form the set $\cP_2\otimes\cP_2$.
 \qed

 The following result was communicated to me by J. Rosenberg.
 Recall that $P\in \cP_n$ is called a \emph{cyclic} permutation
 if $\sum_{i=1}^n P^{i}$ is a matrix whose all entries are
 equal to $1$.

 \begin{lemma}\label{nnexam} Let $P,Q\in \cP_n$ be cyclic
 permutations.  Then the block matrix
 $D=\frac{1}{n}[P^{i}Q^{j}]_{i,j=1}^n$ belongs to $\Phi_{n,n}$.
 If $P\ne Q^i$ for $i=1,\ldots,n-1$
 then $D\not\in\Psi_{n,n}$.  In particular
 $\Psi_{n,n}\subsetneqq \Phi_{n,n}$ for $n\ge 4$.
 For $n=3$ each $D$ of the above form is in $\Psi_{3,3}$.

 \end{lemma}

 \proof  Since $P^i,Q^j\in \Omega_n$  it
 follows that $P^iQ^j\in \Omega_n$ for $i,j=1,\ldots,n$.
 Hence the conditions (\ref{ijcond}) and (\ref{dscon}) are
 satisfied.  It is left to show the conditions (\ref{klcond}).
 Denote $A^i=[a_{kp}^{(i)}]_{k,p=1}^n,
 B^j=[b^{(j)}_{pl}]_{p,l=1}^n\in \Omega_n$.  View
 $D=[c_{(i,k)(j,l)}]$.  Then
 \begin{equation}\label{aibjfor}
 c_{(i,k)(j,l)}=\frac{1}{n}\sum_{p=1}^n
 a^{(i)}_{kp} b_{pl}^{(j)},\quad i,j,k,l=1,\ldots,n.
 \end{equation}
 Since $\sum_{j=1}^n b^{(j)}_{pl}=1$ for $p,l=1,\ldots,n$
 and $A^i\in \Omega_n$ we obtain
 $\frac{1}{n}\sum_{j=1} c_{(i,k)(j,l)}= \frac{1}{n}\sum_{p=1}^n
 a_{kp}^{(i)}=\frac{1}{n}$.  In a similar way we deduce that
 $\sum_{j=1}^n c_{(j,k)(i,l)}=\frac{1}{n}$.  So $D\in
 \Phi_{n,n}$.

 Suppose that $D\in \Psi_{n,n}$.  Observe that $P^nQ^n=I_n
 I_n=I_n$.  Assume  $D$ as a convex
 combination of some extreme points $U\otimes V\in \cP_n\otimes \cP_n$
 with positive coefficients.  Express
 $U\otimes V$ as a block matrix $[(U\otimes
 V)_{ij}]_{i,j=1}^n$.  Suppose furthermore that $(U\otimes
 V)_{nn}\ne 0_{n\times n}$.  Then $V=I_n$.  Hence there exists
 $j\in \an{n-1}$ such $PQ^j=I$, i.e $P=Q^{n-j}$.  If $P$ is not
 a power of $Q$ we deduce that $D\not\in \Psi_{n,n}$.
 For $n\ge 4$ it is easy to construct such two permutations.
 For example, let $P$ and $Q$ are represented by the cycles
 $$1\rightarrow 3 \rightarrow 2\rightarrow 4\rightarrow\ldots
 \rightarrow n\rightarrow 1,\;
 1\rightarrow 2 \rightarrow 3\rightarrow 4\rightarrow\ldots
 \rightarrow n\rightarrow 1.$$

 If $n=3$ then one has only two cycles $R$ and $R^2$.
 A straightforward calculation show that if $P,Q\in\{R,R^2\}$
 the $D\in \Psi_{3,3}$.  \qed

 Note that the system (\ref{dscon}) has $2mn-1$ linear independent
 equations.  Since any permutation matrix is an
 extreme point in $\Omega_{mn}$ we deduce.

 \begin{corol}\label{numbeqphinn}  The convex set $\Phi_{m,n}\subset \R_+^{mn\times mn}$
 is given by at most $2((n-1)m^2+(m-1)n^2+mn)-1$ linear equations.
 It contains all the extreme points $\cP_m\otimes\cP_n$ of
 $\Psi_{m,n}$.
 \end{corol}

 It is interesting to understand the structure of the set $\Phi_{m,n}$
 and to characterize it extreme points.  It is easy to
 characterize the following larger set.

 \begin{lemma}\label{chatetmn}  Let $\Theta_{m,n}$ be the
 convex set of $mn\times mn$ nonnegative
 matrices characterized by $2mn+(2n-2)m^2$ linear equations of the
 following form.  View $C\in \R^{mn\times mn}$ as a matrix with
 entries $c_{(i,k)(j,l)}$ where
 $i,j=1,\ldots,m,\;k,l=1,\ldots,n$.  Then $C\in \R_+^{mn\times
 mn}$ belongs to $\Theta_{m,n}$ if the
 equalities (\ref{dscon}) and (\ref{ijcond}) hold.
 Then
 $\Phi_{m,n}\subset\Theta_{m,n}\subset\Omega_{mn}$.
 Furthermore, any $C=[C_{ij}]_{i,j=1}^m\in \Theta_{m,n}$ is of the following form
 \begin{equation}
 C_{ij}=a_{ij}D_{ij}, \;D_{ij}\in \Omega_n, \;i,j=1,\ldots,m,\;
 A=[a_{ij}]_{i,j=1}^m\in \Omega_m.
 \end{equation}
 In particular, the extreme points of $\Theta_{m,n}$ are of the
 the above form where $A\in\cP_m, D_{ij}\in \cP_n$ for
 $i,j=1,\ldots,n$.
 \end{lemma}

 \proof
 Observe first that $C$ in the block from
 $C=[C_{ij}],\;C_{ij}\in \R^{n\times n}$ where $C_{ij}\in
 \R_+^{n\times n}$.  Conditions  (\ref{ijcond}) equivalent
 to the assumptions that $C_{ij}\in \Lambda_n$.  Hence
 $C_{ij}=f_{ij} D_{ij}$ for some $D_{ij}\in \Omega_n$ and
 $f_{ij}\ge 0$.  If $f_{ij}=0$ we can choose any  $D_{ij}\in \Omega_n$.
 If $f_{ij}>0$ then $D_{ij}$ is a unique doubly stochastic
 matrix.  Let $F=[f_{ij}]\in \R^{m\times m}$.  Then the
 conditions (\ref{dscon}) are equivalent to the condition that
 $F\in \Omega_m$.  Thus the conditions (\ref{ijcond}) and
 (\ref{dscon}) are equivalent to the statement that $C=[f_{ij}
 D_{ij}]$ where each $D_{ij}\in \Omega_n$ and $F=[f_{ij}]\in
 \Omega_m$.

 Since the extreme points of $\Omega_n$ are $\cP_n$ we
 deduce that any extreme point of $\Theta_{m,n}$ is of the block
 form $C=[f_{ij}P_{ij}]$ where each $P_{ij}\in \cP_n$.
 Since the extreme points of $\Omega_m$ are $\cP_m$ it follows
 that the extreme points of $\Theta_{m,n}$ are of the form
 $E=[E_{ij}]$ satisfying the following conditions.  There
 exists a permutation $\sigma:\an{m}\to\an{m}$ such that
 $E_{i\sigma(i)}\in \cP_n$ for $i=1,\ldots,m$ and
 $E_{ij}=0_{m\times m}$ otherwise.
 \qed

 \section{Permutational similarity and equivalence of matrices}
 For $A\in\R^{n\times n}$ denote by $\tr A$ the \emph{trace} of
 $A$.  Recall that $\an{A,B}$, the standard inner product on
 $\R^{n\times n}$, is given by $\tr AB\trans$.

% Denote by $\rS_n\subset \R^{n\times n}$ the real vector space
% of symmetric matrices.
 We say that $A,B\in \R^{n\times n}$ are permutationally
 similar, and denote it by $A\sim B$ if $B=PAP\trans$.
 Clearly, if $A\sim B$ then $A$ and $B$ have the same characteristic polynomial,
 i.e. $\det (xI_n -A)=\det (xI_n-B)$.  In what follows we need
 the following three lemmas.  The proof of the first two
 straightforward and is left to the reader.

 \begin{lemma}\label{nescnpersim}  Let $A=[a_{ij}],B=[b_{ij}]\in \R^{n\times n}$.  Assume
 $A\sim B$.  Then the following conditions hold.
 \begin{eqnarray}
 \label{diagel}
 P(a_{11},\ldots,a_{nn})\trans=(b_{11},\ldots,b_{nn})\trans
 \textrm{ for some } P\in\cP_n,\\
 \label{ofdiagel}
 R(a_{12},\ldots,a_{1n},a_{21},a_{23},\ldots,a_{2n},\ldots,a_{n1},\ldots,a_{n(n-1)})\trans=\\
 (b_{12},\ldots,b_{1n},b_{21},b_{23},\ldots,b_{2n},\ldots,b_{n1},\ldots,b_{n(n-1)})\trans
  \textrm{ for some } R\in\cP_{n^2-n}.
  \nonumber
 \end{eqnarray}
 \end{lemma}

 \begin{lemma}\label{traceq} Assume that $A,B\in\R^{n\times n}$ satisfy
 the conditions (\ref{diagel}) and (\ref{ofdiagel}).  Then
 $\tr (A+tI_n)(A+tI_n)\trans=\tr(B+tI_n)(B+tI_n)\trans$ for each $t\in \R$.
 \end{lemma}

 \begin{lemma}\label{eqfrmpers}  Let $A=[a_{ij}],B=[b_{ij}]\in \R^{n\times n}$
 satisfy the conditions (\ref{diagel}) and (\ref{ofdiagel}).
 Fix $t\in R$ such that $t\ne a_{ij}-a_{kk}$ for each $i,j,k\in
 \an{n}$ such that $i\ne j$.
 Then the following conditions are equivalent.
 \begin{enumerate}
 \item\label{eqfrmpers1} $A\sim B$.
 \item\label{eqfrmpers2} $B+tI_n=P(A+tI_n)Q\trans$ for some $P,Q\in
 \cP_n$.

 \end{enumerate}

 \end{lemma}

 \proof  Suppose that
 \emph{2} holds.  Hence there exists two permutations
 $\sigma,\eta:\an{n}\to\an{n}$ such that

 $$b_{ij}+t\delta_{ij}=a_{\sigma(i)\eta(j)}+t\delta_{\sigma(i)\eta(j)}
 \quad \textrm{ for all } i,j\in \an{n}.$$
 Assume that $\sigma\ne \eta$.  Then there exists $i\ne j\in
 \an{n}$ such that $\sigma(i)=\eta(j)=k$.  Hence
 $b_{ij}=a_{kk}+t$.  The condition (\ref{ofdiagel}) implies
 that $b_{ij}=a_{i_1j_1}$ for some $i_1\ne j_1\in \an{n}$.
 So $t=a_{i_1j_1}-a_{kk}$, which contradicts the assumptions of
 the lemma.  Hence $\sigma=\eta$ which is equivalent to $P=Q$.
 Thus

 $$B+tI_n=P(A+tI_n)P\trans=PAP\trans +tI_n \Rightarrow
 B=PAP\trans.$$
 Reverse the implication in the above statement to deduce \emph{2} from \emph{1}.
 \qed

 We recall standard facts from linear algebra.
 \begin{lemma}\label{replinop}
 Let $X=[x_{lj}]_{l,j=1}^{n,m}=[\x_1\;\x_2\ldots\x_m]\in\R^{n\times m}$, where
 $\x_1,\ldots,\x_m\in\R^n$ are the $m$ columns of $X$.  Denote
 by $\hat X\in \R^{mn}$ the column vector composed of the
 columns of $X$, i.e. $(\hat
 X)\trans=(\x_1\trans,\x_2\trans,\ldots,\x_m\trans)$.  Let
 $A=[a_{ij}]\in \R^{m\times m}, B=[b_{kl}]\in \R^{n\times n}$.  Consider the
 linear transformation of $\R^{m\times n}$ to itself given by
 $X\mapsto BXA\trans=[(BXA\trans)_{ki}]_{k,i=1}^{n,m}$:
 \begin{equation}\label{replinopmat}
 (BXA\trans)_{ki}=\sum_{j,l=1}^{m,n} a_{ij}b_{kl} x_{lj}, \quad
 k=1,\ldots,n,\;i=1,\ldots,m.
 \end{equation}
 Then this linear transformation is
 represented by the Kronecker product $A\otimes B$.  That is,
 \begin{equation}\label{replinop1}
 \widehat{BXA\trans}= (A\otimes B)\hat X \quad \textrm{ for all }
 X\in \R^{n\times m}.
 \end{equation}

 \end{lemma}

 \proof Observe first that $BX=[B\x_1\;B\x_2\ldots B\x_m]$.
 This shows (\ref{replinop1}) in the case $A=I_m$.  Consider
 now the case $B=I_n$.  A straightforward calculation shows that
 $(A\otimes I_n)\hat X= \widehat{XA\trans}$.  Since $A\otimes
 B=(A\otimes I_n) (I_m\otimes B)$ we deduce the equality
 (\ref{replinop1}).  \qed

 \begin{theo}\label{mthmpersim}  Let $A=[a_{ij}],B=[b_{ij}]\in \R^{n\times n}$.
 The following conditions are equivalent.
 \begin{enumerate}

 \item\label{mthmpersim1}  $A\sim B$.

 \item\label{mthmpersim2}  The following conditions hold.
 \begin{enumerate}
 \item \label{mthmpersim3}  The conditions (\ref{diagel}) and
 (\ref{ofdiagel}) hold.
 \item  \label{mthmpersim4}  Fix $t\in R$ such that $t\ne a_{ij}-a_{kk}$ for each $i,j,k\in
 \an{n}$ such that $i\ne j$.
 Then there exists
 $Z\in \Psi_{n,n}$  satisfying
 $Z\widehat {(A+t  I_n)}= \widehat{B+tI_n}$.

 \end{enumerate}

 \end{enumerate}

 \end{theo}

 \proof Assume \emph{1}.  So $B+tI_n=P(A+tI_n)P\trans$ for some $P\in \cP_n$ and each $t\in \R$.
 Use Lemma \ref{replinop} to deduce that $(P\otimes P) \widehat {(A+t  I_n)}= \widehat{B+tI_n}$.
 Hence the condition \emph{2b} holds.
 Lemma \ref{nescnpersim} yields the conditions (\ref{diagel}) and
 (\ref{ofdiagel}).

 Assume \emph{2}.  Use Lemma \ref{traceq} yields that $\tr
 (A+tI_n)(A+tI_n)\trans = \tr (B+tI_n)(B+tI_n)\trans$.
 We claim that
 \begin{equation}\label{maxchar}
 \max_{P,Q\in \cP_n} \tr P(A+tI_n)Q\trans (B+t I_n)\trans=
 \max_{Y\in \Psi_{n,n}} (\widehat{B+t I_n})\trans Y
 (\widehat{A+tI_n}).
 \end{equation}
 To find the maximum on the right-hand side it is enough to
 restrict the maximum on the right-hand side to the extreme
 points of $\Psi_{n,n}$.  Lemma \ref{conhtproddsm} yields that
 the extreme points of $\Psi_{n,n}$ are $\cP_n\otimes \cP_n$.
 Let $Y=Q\otimes P\in \cP_n\otimes \cP_n$.  (\ref{replinop1})
 yields that
 $$(\widehat{B+t I_n})\trans Y (\widehat{A+tI_n})=\tr
 P(A+tI_n)Q\trans (B+tI_n)\trans.$$
 Compare the above expression with the left-hand side of
 (\ref{maxchar}) to deduce the equality in (\ref{maxchar}).

 Assume that the maximum in the left-hand side of (\ref{maxchar})
 is achieved for $P_*,Q_*\in \cP_n$.  Use Cauchy-Schwarz
 inequality to deduce that
 \begin{eqnarray*}
 \tr P_*(A+tI_n)Q_*\trans (B+t I_n)\trans\le\\
 ((\tr P_*(A+tI_n)(A+tI_n)\trans P_*\trans)
 \tr (B+tI_n)(B+tI_n)\trans)^{\frac{1}{2}}=
 \tr (B+tI_n)(B+tI_n)\trans.
 \end{eqnarray*}
 Equality holds if and only if $B+tI_n=P_*(A+tI_n)Q_*\trans$.
 The assumption \emph{2b} yields the opposite inequality
 \begin{eqnarray*}
 \tr (B+tI_n)(B+tI_n)\trans=(\widehat{B+t I_n})\trans Z (\widehat{A+tI_n})
 \le\\
 \max_{Y\in \Psi_{n,n}} (\widehat{B+t I_n})\trans Y
 (\widehat{A+tI_n})=\tr P_*(A+tI_n)Q_*\trans (B+t I_n)\trans.
 \end{eqnarray*}

 Hence $B+tI_n=P_*(A+tI_n)Q_*\trans$.  Lemma \ref{eqfrmpers}
 implies that $A\sim B$.
 \qed

 The proof of the above theorem yields.

 \begin{corol}\label{extptsxab}  Assume that the conditions
 \emph{2} of Theorem \ref{mthmpersim} holds.  Let
 $\Psi_{n,n}(A,B)$ be the set of all $Z\in \Psi_{n,n}$
 satisfying the condition $Z(\widehat {A+tI_n})=\widehat
 {B+tI_n}$.  Then all the extreme points of this compact convex
 set are of the form $P\otimes P\in \cP_n\otimes\cP_n$ where
 $PAP\trans =B$.
 \end{corol}

 $A,B\in \R^{n\times n}$ are called \emph{permutationally
 equivalent}, denoted as $A\approx B$, if $B=PAQ\trans$ for some $P\in\cP_n, Q\in\cP_m$.
 The arguments of the proof of Theorem \ref{mthmpersim} yield.
 \begin{theo}\label{mthmpereq}  Let $A,B\in \R^{n\times m}$.
 The following conditions are equivalent.
 \begin{enumerate}

 \item\label{mthmpereq1}  $A\approx B$.

 \item\label{mthmpereq2}

 $\tr A A\trans=\tr B B\trans$ and
 there exists
 $Z\in \Psi_{m,n}$  satisfying
 $Z\widehat {A}= \widehat{B}$.
 That is, view the entries
 of $Z$ as $z_{(i,k),(j,l)}$ where $i,j\in\an{m},k,l\in\an{n}$.  Then
 these $m^2n^2$ nonnegative variables satisfy $2((n-1)m^2+(m-1)n^2+mn)$
 conditions (\ref{dscon}-\ref{ijcond}) and the $mn$
 conditions:
 \begin{equation}\label{mn2cond}
 \sum_{j,l=1}^{m,n}z_{(i,k)(j,l)}a_{lj}=b_{ki}
 \textrm{ for } k=1,\ldots,n,\;i=1,\ldots,m.
 \end{equation}

 \end{enumerate}

 \end{theo}

 \begin{corol}\label{extptsxab1}  Assume that the conditions
 \emph{2} of Theorem \ref{mthmpereq} holds.  Let
 $\Psi_{m,n}(A,B)$ be the set of all $Z\in \Psi_{m,n}$
 satisfying the condition $Z\widehat {A}=\widehat
 {B}$.  Then all the extreme points of this compact convex
 set are of the form $Q\otimes P\in \cP_m\otimes\cP_n$ where
 $PAQ\trans =B$.
 \end{corol}

 \section{GIP and SGIP}
 \subsection{Graph isomorphisms}

 \begin{theo}\label{graphisudg}
 Assume that $\Psi_{n,n}$ is characterized by $f(n)$
 number of linear equalities and inequalities.
 Then isomorphism
 of two simple undirected graphs $G_1=(V,E_1)$, $G_2=(V,E_2)$ where $\#V=n$
 is decidable in polynomial time in $\max(f(n),n)$.
 \end{theo}

 \proof  Let $A,B\in \{0,1\}^{n\times n}$ be the adjacency
 matrices of $G_1,G_2$ respectively.  Recall that $A,B$ are
 symmetric and have zero diagonal.
 $G_1$ and $G_2$ are isomorphic if and only if $A\sim B$.
 It is left to show that
 the conditions \emph{2} of Theorem \ref{mthmpersim} can be
 verified in polynomial time in $\max(f(n),n)$.  \emph{2a} means that $G_1$ and
 $G_2$ have the same degree sequence.  This requires at most
 $4n^2$ computations.  Assume that \emph{2a} holds.
 Note that $t=2$ satisfies the first part of the condition \emph{2b}.
 The existence of $Z\in
 \Psi_{n,n}$ satisfying $Z (\widehat{A+2I_n})=\widehat{B+2I_n}$
 is equivalent to the solvability of $f(n)+n^2$
 linear equations and inequalities in
 $n^4$ nonnegative variables.
 The ellipsoid method \cite{Kha, Kar} yields that the existence
 or nonexistence of such $X\in \Psi_{n,n}$ is decidable in
 polynomial time in $\max(f(n),n)$. \qed

 \begin{theo}\label{graphisdg}
 Assume that $\Psi_{n,n}$ is characterized by $f(n)$
 number of linear equalities and inequalities.
 Then the isomorphism
 of two simple directed graphs $G_1=(V,E_1)$, $G_2=(V,E_2)$, (self-loops allowed),
 where $\#V=n$
 is decidable in polynomial time in $\max(f(n),n)$.
 \end{theo}
 \proof
 Let $A,B\in \{0,1\}^{n\times n}$ be the adjacency
 matrices of $G_1,G_2$ respectively.  Apply part \emph{2}
 of Theorem \ref{mthmpersim} with $t=2$ to deduce the theorem.
 \qed

 The application of part \emph{2}
 of Theorem \ref{mthmpersim} yields.

 \begin{theo}\label{polpersimmat}
 Assume that $\Psi_{n,n}$ is characterized by $f(n)$
 number of linear equalities and inequalities.
 Let $A,B\in \R^{n\times n}$.
 Then permutational similarity of $A$ and $B$ is decidable in
 polynomial time in $\max(f(n),n)$ and the entries of $A$ and $B$.
 \end{theo}

 Let $G=(V_1\cup V_2,E)$ be an undirected simple bipartite
 graph with the set of vertices divided to two classes
 $V_1,V_2$ such that $E\subset V_1\times V_2$.
 Assume that $\#V_1=n, \#V_2=m$ and identify $V_1,V_2$
 with $\an{n},\an{m}$ respectively.  Then $G$ is represented by
 the incidence matrix $A=[a_{ij}]\in \{0,1\}^{n\times m}$ where
 $a_{ij}=1$ if and only if the vertices $i\in\an{n}, j\in
 \an{m}$ are connected by an edge in $E$. Let $H=(V_1\cup
 V_2,F)$ be another bipartite graph with the incidence
 matrix $B\in \{0,1\}^{n\times m}$.  If $m\ne n$ then $G$ and
 $H$ are isomorphic if and only if $A\approx B$.  If $m=n$
 $G$ and $H$ are isomorphic if and only if either $A\approx B$ or
 $A\approx B\trans$.  Theorem \ref{mthmpereq} yields.

 \begin{theo}\label{graphisbip}
 Assume that $\Psi_{m,n}$ is characterized by a $g(m,n)$
 number of linear equalities and inequalities.
 The isomorphism
 of two simple undirected bipartite graphs $G_1=(V_1\cup V_2,E_1)$,
 $G_2=(V_1\cup V_2,E_2)$ where $\#V_1=n,V_2=m$
 is decidable in polynomial time in $\max(g(m,n),n+m)$.
 \end{theo}

 \begin{theo}\label{polpereqmat}
 Assume that $\Psi_{m,n}$ is characterized by $g(m,n)$
 number of linear equalities and inequalities.
 Let $A,B\in \R^{n\times m}$.
 Then permutational equivalence of $A$ and $B$ is decidable in
 polynomial time in $\max(g(m,n),n+m)$ and the entries of $A$ and $B$.
 \end{theo}

 We now remark that if we replace in Theorem \ref{mthmpersim}
 and Theorem \ref{mthmpereq} the sets $\Psi_{n,n}$ and
 $\Psi_{m,n}$ by the sets $\Phi_{n,n}$ and
 $\Phi_{m,n}$ respectively, we will obtain necessary conditions
 for permutational similarity and equivalence, which can be
 verified in polynomial time.

 \subsection{Subgraph isomorphism}
 \begin{theo}\label{subgraphisudg}
 Assume that $\Psi_{n,n}$ is characterized by $f(n)$
 number of linear equalities and inequalities.
 Let $G_3=(W,E_3), G_2=(V,E_2)$ be two simple undirected
 graphs, where $\#W=m\le \#V=n$.
 Then the problem of determining if $G_3$ is isomorphic to a
 subgraph of $G_2$
 is decidable in polynomial time in $\max(f(n),n)$.
 \end{theo}

 \proof
 Add $n-m$ isolated vertices to $G_3$ to obtain the
 graph $\tilde G_3$ on $n$ vertices.
 Let $C,B\in \{0,1\}^{n\times n}$ be the adjacency
 matrices of $\hat G_3,G_2$ respectively.
 We claim that $G_3$ is isomorphic to a subgraph of $G_2$
 if and only if
 \begin{equation}\label{subgisoin}
 Z (\widehat{C+2^{n^2}I_n})\le
 \widehat{B+2^{n^2}I_n} \textrm{ for some } Z\in
 \Psi_{n,n}.
 \end{equation}

 Assume first that $G_3$ is isomorphic to a subgraph of $G_2$.
 This is equivalent to the statement that $PCP\trans\le B$ for some
 $P\in \cP_n$.  (That is in each place where $PCP\trans$ has
 entry $1$, then $B$ has entry $1$ at the same place.)
 As $PP\trans =I$ we deduce that (\ref{subgisoin}) holds for
 $Z=P\otimes P$.

 Assume that (\ref{subgisoin}) is satisfied.  Let
 $$Z=\sum_{P,Q\in \cP_n} w(P,Q)P\otimes Q,\; w(P,Q)\ge 0 \textrm{ for each }
 P,Q\in \cP_n \textrm{ and } \sum_{P,Q\in\cP_n} w(P,Q)=1.$$
 Hence there exists $P_*,Q_*\in\cP_n$ such that $w(P_*,Q_*)\ge
 \frac{1}{(n!)^2}$.  (\ref{subgisoin}) yields that
 $$\frac{1}{(n!)^2} Q_*(C+2^{n^2} I_n) P_*\trans \le
 B+2^{n^2}I_n.$$
 Since $n= 2^{n-1}$ for $n=1,2$ and $n<2^{n-1}$ for $2<n$ it follows
 that $n!<2^{\frac{n(n-1)}{2}}$ for $n>2$.  Hence $(n!)^2<
 2^{n^2}$ for $n\ge 1$.  Since all offdiagonal elements of $B$
 are at most $1$ it follows that $P_*=Q_*$. Hence
 $P_*CP_*\trans \le (n!)^2 B$.  Thus if $P_*CP_*\trans$ has $1$
 in the place $(i,j)$ then $B$ can not have zero in the place
 $(i,j)$.  That is $B$ has $1$ in the place $(i,j)$.  Therefore
 $G_3$ is isomorphic to a subgraph of $G_2$.  \qed

\end{document}